\documentclass[twocolumn,aps,pre,showpacs,preprintnumbers,groupedaddress,amsmath,amssymb,floatfix]{revtex4-1}
\usepackage{graphicx,CJK,times}
\usepackage{color}

\begin{document}

\begin{CJK*}{UTF8}{}
\title{Enhanced tunneling conductivity induced by gelation of attractive colloids} 
 
\author{Biagio Nigro}
\affiliation{LPM, Ecole Polytechnique F\'ed\'erale de Lausanne (EPFL), CH-1015
Lausanne, Switzerland}  
\author{Claudio Grimaldi}\email{claudio.grimaldi@epfl.ch}
\altaffiliation{present address: Laboratory of Physics of Complex Matter,
Ecole Polytechnique F\'ed\'erale de Lausanne (EPFL), CH-1015
Lausanne, Switzerland}
\affiliation{LPM, Ecole Polytechnique F\'ed\'erale de Lausanne (EPFL), CH-1015
Lausanne, Switzerland} 
\author{Peter Ryser}
\affiliation{LPM, Ecole Polytechnique F\'ed\'erale de Lausanne (EPFL), CH-1015
Lausanne, Switzerland} 

\author{Francesco Varrato} 
\affiliation{Institute of Theoretical Physics, Ecole Polytechnique F\'ed\'erale
de Lausanne (EPFL), CH-1015 Lausanne, Switzerland}
\author{Giuseppe Foffi}
\affiliation{Institute of Theoretical Physics, Ecole Polytechnique F\'ed\'erale
de Lausanne (EPFL), CH-1015 Lausanne, Switzerland}
\affiliation{Laboratoire de Physique de Solides, UMR 8502,
Bat. 510, Universite Paris-Sud, F-91405 Orsay, France}

\author{Peter J.~Lu (\CJKfamily{bsmi}陸述義)}
\affiliation{Department of Physics and SEAS, Harvard University, Cambridge,
Massachusetts 02138, USA}

\begin{abstract}
We show that the formation of a gel by conducting colloidal particles leads to a
dramatic enhancement in bulk conductivity, due to inter-particle electron
tunneling, combining predictions from molecular dynamics simulations with
structural measurements in an experimental colloid system. Our results show how
colloidal gelation can be used as a general route to huge enhancements of
conductivity, and suggest a feasible way for developing cheap materials with
novel properties and low metal content.
\end{abstract}
\maketitle
\end{CJK*}

\section{introduction}
\label{intro}
Attractive forces between colloidal particles, ranging in size from nanometers
to microns, can be finely controlled in the laboratory through surface
functionalization or addition of depletants, leading to a wide range of
phases and dynamical behaviors~\cite{Poon1998, Lekke2002, Sciortino2002,
Frenkel2006, Lekke2011}. In particular, when the inter-particle attraction range
$\lambda$ is much smaller than the particle diameter $D$, the dynamical arrest
of phase separation leads to the formation of colloidal gels---arrested,
space-spanning structures---even at low particle volume fractions
$\phi$~\cite{Manley2004,Lu2008}. Gels form when particles are quenched into the
liquid-gas phase-separation region of the phase diagram, and spinodal
decomposition arrests~\cite{Foffi2005,Lu2008,Zaccarelli2008}; they have been
observed experimentally in colloids and
proteins~\cite{Lu2006,Lu2008,Poon2002,Cipelletti2005,Cardinaux2007,Zaccarelli2008},
and in computer simulations\cite{Soga1998, Foffi2005, FoffiJCP, Gado2010}. The
structure of these gels depend on $\phi$, $\lambda$ and the strength of the
attraction~\cite{Lu2006}, and can form sparse, ramified semi-solid structures.

For dispersions of conducting particles, tuning the inter-particle attractions
can alter the bulk electrical properties of the
system~\cite{Vigolo2005,Kyrylyuk2011}, for example enhancing electrical
conductivity in colloidal fluids~\cite{Schilling2007} by lowering the mean
distance required for electrons to tunnel between particles~\cite{Nigro2012}. If
this enhanced tunneling were to persist or be amplified in the deeply-quenched
region as well, this might provide a new route to designing materials with novel
combinations of electrical and mechanical properties, such as new
highly-conductive semi-solid materials which could have important practical
applications in energy storage and transport.

In this Article, we explore how the gelation of conducting colloidal particles
affects overall electrical conductivity, where electron tunneling between
particles is the principal mechanism of electrical conductivity.
We find using molecular dynamics computer simulations that the tunneling
conductivity $\sigma$ in the arrested gel phase depends only weakly on $\phi$,
and remains relatively high even for $\phi$ as low as $3$\%.
In this regime, we also find that $\sigma$ is only moderately affected by
varying $\xi/D$ by as much as one order of magnitude, where $\xi$ is the
characteristic tunneling decay length. In addition, we perform the same analysis
for gel structures formed in an experimental colloidal system with short-ranged,
attractive depletion interactions; we find a similar shortening of the relevant
tunneling distances as the system evolves towards the arrested gel state. Our
results demonstrate that conduction via tunneling in gels of conducting
colloidal particles can occur using realistic assumptions of microscopic
parameters, opening up the possibility of creating new, lightweight, highly
conductive materials with novel mechanical and electrical properties.

\section{model and simulations}
\label{model}
Our simulations comprise a colloidal system of $N$ conducting monodisperse
spherical particles dispersed in an continuous insulating medium, with volume
fraction $\phi=\pi\rho D^3/6$, where $D$ is the sphere diameter, $\rho=N/L^3$ is
the number density and $L$ is the box size. We assume that the conductance
between any two particles $i$ and $j$ is dominated by electron tunneling
processes, with conductance $g(\delta_{ij})$:
\begin{equation}
\label{gij}
g(\delta_{ij})=g_0\exp\!\left(-\frac{2\delta_{ij}}{\xi}\right),
\end{equation}
where $\delta_{ij} \equiv r_{ij} - D$ is the closest distance between particle
surfaces, $r_{ij}$ is the center-to-center distance between particles, $\xi$ is
the tunneling decay length, and $g_0$ is a prefactor that we define so that the
conductance between two touching colloids is $g(0) \equiv 1$.
The potential barrier separating conducting and insulating phases determines
$\xi$, which typically ranges from a fraction of a nanometer to a few
nanometers~\cite{xivalues}. Consequently, $\xi/D \lesssim 0.1$ for particles
larger than a few tens of nanometers, so that, consistently with
Eq.~\eqref{gij}, charging and Coulomb interaction effects on electron transfer
can be safely neglected at room temperature.

For colloidal systems with short-ranged attractions, $\lambda/D \lesssim 0.05$,
thermodynamic properties at a given $\phi$ depend not on the specific shape of
the potential $u(r)$, but only on its integral, expressed as a reduced second
virial coefficient $B_2^*=(3/D^3) \int  [1-e^{-u(r)/T}]  r^2 dr$, where $T$ is
the temperature and $k_B \equiv
1$~\cite{Noro2000,Foffi2006,Malijevsky2006,Lu2008}. In particular, short-ranged
attractive colloidal spheres are in an equilibrium fluid phase for
$B_2^{*}\gtrsim B_2^{*c}$, where $B_2^{*c}\simeq -1.2$ is the critical value at
the critical point of the gas-liquid phase
separation~\cite{Miller2003,Largo2008}. Because all short-range potential shapes
yield the same thermodynamic behavior~\cite{Noro2000,Lu2008}, we select a
square-well (SW) model of the interaction of the following form:
\begin{equation}
\label{SWmodel}
u(\delta_{ij})=\left\{
\begin{array}{ll}
\infty & \delta_{ij}\leq 0 \\
-u_0 & 0< \delta_{ij}\leq \lambda D \\
0 & \delta_{ij}>\lambda D
\end{array}\right.
\end{equation}
where $\lambda \ll 1$ and $u_0>0$ are, respectively, the dimensionless potential
range and depth. As in Eq.~\eqref{gij}, $\delta_{ij}$ denotes the closest
distance between surfaces of particle pairs. The scaled virial coefficient of
the potential in Eq.~\eqref{SWmodel} can be expressed as $B_2^* \equiv 1-1/4\tau$,
where $\tau^{-1}\equiv 4[(1+\lambda)^3-1][\exp(u_0/T)-1]$ is the Baxter
stickiness parameter~\cite{Baxter1968}. Consequently, a homogeneous SW fluid
exists when $\tau$ is greater than the critical value $\tau_c\sim
0.11$~\cite{Miller2003,Largo2008}.

In this regime, SW fluids of conducting particles display enhanced conductivity
$\sigma$, relative to the hard-sphere case, as $\tau$ is
decreased~\cite{Nigro2012}.
The inter-particle attraction enhances conductivity by drawing the particles
closer together: the population of particles with separations lower than
$\lambda D$ increases, thereby promoting short-length tunneling processes, which
result in larger $g(\delta_{ij})$.
Specifically, in attractive colloidal fluids where $\lambda\rightarrow 0$,
$\tau=0.2$, and $\xi/D=0.01$, $\sigma$ is relatively large and depends only
weakly on $\phi$ for $\phi \gtrsim 0.2$.\cite{Nigro2012}

When $B_2^*\leq B_2^{*c}$  (i.e., $\tau \leq \tau_c$), short-ranged attractive
colloids undergo phase separation and can arrest to form gels: spanning
structures that may sustain shear stresses even at low $\phi$~\cite{Lu2008,
Foffi2005, FoffiJCP, Gado2010}. In these gel configurations, larger tunneling
conductivities might be expected relative to the fluid phase at the same $\phi$,
as the mean separation between the particles forming the gel network falls below
$\lambda D$; however, knowledge of potential range alone is insufficient to
predict the conductivity level of the system, as previously shown for SW fluids
of conducting colloids~\cite{Nigro2012}. Instead, the full interplay between
tunneling, attraction and structure must be considered.

\section{molecular dynamics simulations}
\label{simulations}
\begin{figure}
\begin{center}
\includegraphics[scale=0.33,clip=true]{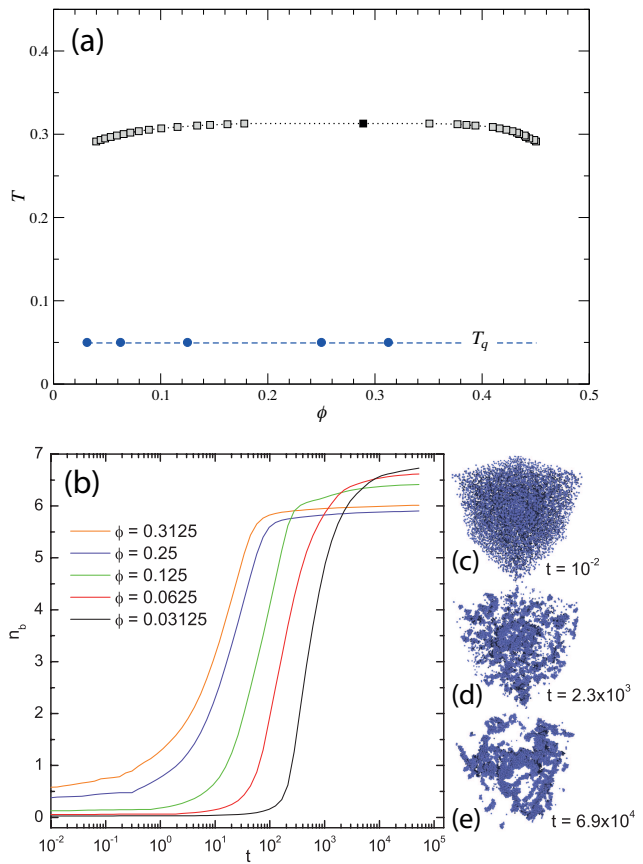}
\caption{
(Color online) (a) Phase diagram for the $\lambda=0.03$ SW system.
The binodal and the critical point are represented by open and closed squares,
respectively. The gel points, obtained by quenching to state points at low
temperature $T_q$, are represented by circles. (b) Average number of bonds $n_b$
for different $\phi$ as a function of $t$, expressed in units of
$D\sqrt{m/u_0}$. Three different stages of aggregation for $\phi=0.03125$ are
shown for (c) $t=0.01$, (d) $t=2.3\times 10 ^3$, and (e) $t=6.9\times 10^4$. The
configuration shown in (e) represents an arrested colloidal gel, where the
particle positions do not display any subsequent time evolution.}\label{fig1}
\end{center}
\end{figure}
\begin{figure*}
\begin{center}
\includegraphics[scale=0.9,clip=true]{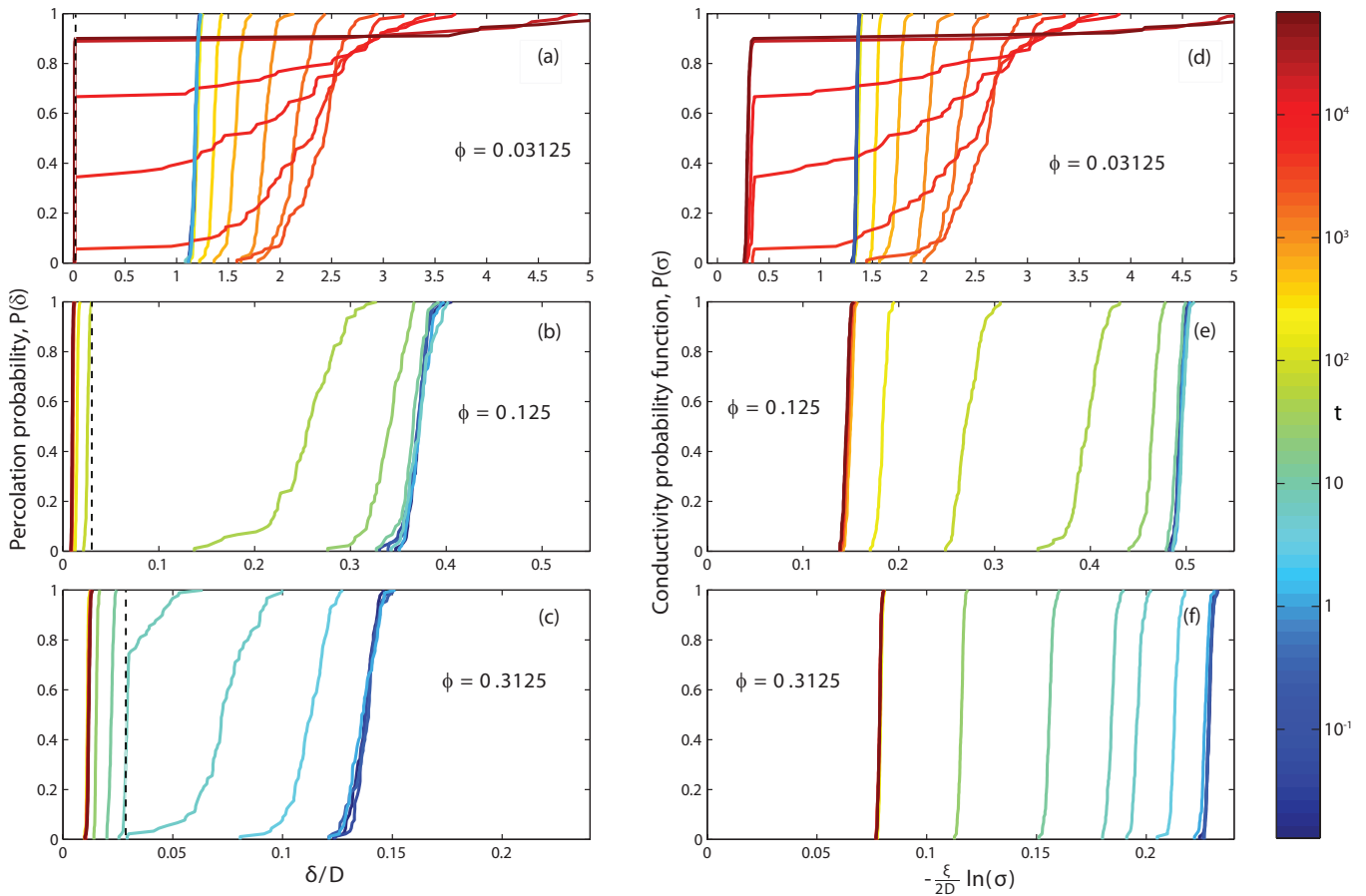}
\caption{
(Color online) Time evolution of percolation probability $P(\delta)$ for the
connectivity distance $\delta / D$ for (a) $\phi=0.03125$, (b) $\phi=0.125$, and
(c) $\phi=0.3125$. The vertical dashed line in (a)-(c) indicates
$\delta/D=\lambda=0.03$. For each $\phi$, each curve represents $P(\delta)$
calculated at a specific time ranging from $t=0.013$ to $t=6.86\times 10^4$,
illustrated with the color (gray) bar at right. (d)-(f): conductivity probability
$P(\sigma)$ as function of $-\frac{\xi}{2D}\ln(\sigma)$ for tunneling decay
length fixed at $\xi/D=0.1$. }
\label{fig2}
\end{center}
\end{figure*}

We generate colloidal gel structures from molecular dynamics (MD) simulations of
$N=10^4$ attractive colloids of mass $m$, square well depth $u_0=1$, and
$\lambda=0.03$, corresponding to a critical temperature $T_c\simeq 0.3$.
We implement Newtonian dynamics via a standard event-driven
algorithm~\cite{Rapaport1997, FoffiJCP}. At $t=0$ we equilibrate initial
configurations at $T=100 \gg T_c$, where these systems behave as hard-sphere
(HS) fluids: $B_2^*\sim 1$. For each selected $\phi$ value we consider $30$
independent realizations. We define two particles as bonded when
$\delta_{ij}\leq \lambda D$, so that the average number of bonds per particle is
$n_\mathrm{b}=-2U/(N u_0)$, where $U/N$ is the potential energy per particle.
For $t>0$, we quench the system to $T_q=0.05 \ll T_c$, corresponding to
$\tau\simeq 5.56\times 10^{-9}$ and $B_2^*\simeq -4.5\times 10^7$; we select
five different packing fractions, ranging from $\phi \simeq 0.03$ to $\phi
\simeq 0.3$, marked with circles in Fig.~\ref{fig1}(a). These
configurations fall well within the two-phase region of the phase diagram;
the full gas-liquid coexistence line at this attraction
range~\cite{Miller2003,Largo2008,Miller2} is marked with squares in the figure.

At short times, the samples appear homogeneously dispersed, as illustrated by a
sample with $\phi=0.03125$ in Fig.~\ref{fig1}(c); this fluid-like initial state,
where $n_\mathrm{b}$ remains at a constant low level, persists longer for lower
$\phi$, as shown in Fig.~\ref{fig1}(b). Following this initial transient state,
concentration fluctuations arising from spinodal decomposition grow rapidly,
marked by a steep rise in $n_\mathrm{b}$, as illustrated in Fig.~\ref{fig1}(d).
Eventually, when the particles become so dense locally as to potentially undergo
an attractive glass transition~\cite{Lu2008}, the structures arrest to form
gels, as illustrated in Fig.~\ref{fig1}(e), and $n_\mathrm{b}$ plateaus at about
$6$; no further evolution is observed.

\section{critical path approximation}
\label{critical}
To explore the effect of gelation on the system conductivity $\sigma$ we first
apply the critical path approximation (CPA) to the network formed by the
tunneling conductances of Eq.~\eqref{gij}~\cite{CPApapers}. When the
$\delta_{ij}$ distances are widely distributed on a length scale of the order of
$\xi$, the CPA provides a robust estimate of the network conductivity through
\begin{equation}
\label{CPA}
\sigma_\mathrm{cpa}=\sigma_0\exp\!\left(-\frac{2\delta_c}{\xi}\right),
\end{equation}
where $\sigma_0$ is a constant prefactor and $\delta_c$ is the shortest
$\delta_{ij}$ such that the subnetwork defined by the bonds satisfying
$\delta_{ij} \leq \delta_c$ forms a percolating
cluster~\cite{Nigro2012,Ambrosetti2010}.
Equation~\eqref{CPA} thus replaces the problem of solving the tunneling network
equations by a simpler one: finding the critical distance $\delta_c$ so that
percolation is established. Depending on the particle concentration and the
potential depth, $\delta_c$ may be larger or lower than the attraction range
$\lambda D$~\cite{Nigro2012}.

\begin{figure}
\begin{center}
\includegraphics[scale=0.4,clip=true]{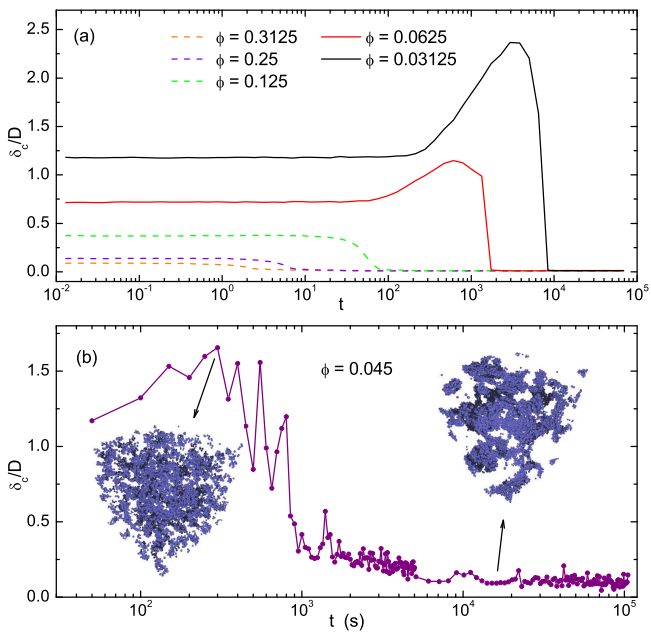}
\caption{
(Color online) (a) Time evolution of the critical connectedness distance
$\delta_c$ for gels simulated at different $\phi$, with $t$ in units of
$D\sqrt{m/u_0}$. (b) Time evolution (in seconds) of the critical distance
extracted from the measured spatial positions of PMMA particles in a
polymer-colloid system~\cite{Lu2008}. For times larger than about $10^4$ s the
system is in an arrested gel state.}\label{fig3}
\end{center}
\end{figure}
To calculate $\delta_c$, we coat each conducting sphere with a concentric
penetrable shell of thickness $\delta_0/2$, for each configuration of the system
at a given volume fraction $\phi$ and time $t$; we consider two spheres as
connected if their penetrable shells overlap (i.e., if $\delta_{ij}\leq
\delta_0$). Using a clustering algorithm~\cite{Nigro2011}, we compute the
minimum value $\delta$ of $\delta_0$ such that a cluster of connected particles
connects two opposite faces of the simulation box. We repeat this procedure
along all three axes of the cubic box, so that a total of $90$ values of
$\delta$ are calculated for each $\phi$ and $t$. By counting the number of
instances for which sample-spanning clusters appear for a given $\delta$, we
construct the percolation probability $P(\delta)$, shown in
Figs.~\ref{fig2}(a)-\ref{fig2}(c) for systems with $\phi=0.03125, 0.125$ and
$0.3125$, respectively.
In each panel of Figs.~\ref{fig2}(a)-\ref{fig2}(c) we plot $P(\delta)$ for
several values of $t$, expressed in units of $D\sqrt{m/u_0}$, ranging from the
onset of the quench at $t=0.013$, to the longest time $t=6.86\times 10^4$, when
the gel phase is fully formed for all $\phi$.

Gelation considerably lowers the values of $\delta$ for which $P(\delta)$
increases from $0$ to $1$, as shown in Fig.~\ref{fig2}(a), indicating that gel
networks percolate at smaller distances than that of the fluid state at the same
$\phi$. In addition, we observe a sudden change in the slope of $P(\delta)$ when
$\delta/D$ crosses $\lambda$, marked with vertical dashed lines, due to the
discontinuity of the SW potential, as seen for $\phi=0.3125$ and $0.03125$ and
already observed in SW fluids~\cite{Nigro2012}.
Furthermore, for $\phi=0.03125$ the percolation transition is not monotonic and,
at large times, $P(\delta)$ reaches the unity only for very large $\delta$; in
these cases, the finite size of the system prevents some realizations of the gel
network with $\phi=0.03125$ from connecting two opposite faces of the box.
We minimize finite-size effects by choosing suitable criteria for extracting
$\delta_c$ from $P(\delta)$; using the criterion $P(\delta_c)=1/2$ to define the
critical distance, we find estimates of $\delta_c$ from $N=10^4$ particles to
differ from the $N\rightarrow \infty$ limit by only a few percent, as shown
in the Appendix.

To understand the dynamics of these systems, we investigate the time evolution
of $\delta_c$ for all $\phi$. At short times, particles are dispersed nearly
homogeneously, so that $\delta_{ij}$, and thus $\delta_c$, decrease strongly as
$\phi$ increases, as shown in Fig.~\ref{fig3}(a) and in agreement with previous
results on SW equilibrium fluids~\cite{Nigro2012}. However, when the system is
arrested at long times, the vast majority of particles forming the spanning gel
structure have separations lower than $\lambda D$, consistent with the data
shown in Fig.~\ref{fig1}(b), where the number of bonds stabilizes at $n_b\approx
6$ for large $t$. Consequently, $\delta_c$ becomes small, about $0.01 D$, and
independent of $\phi$, as shown in Fig.~\ref{fig3}(a). Strikingly, though the
final value of $\delta_c$ is the same for all $\phi$, we observe significant
$\phi$-dependent differences in reaching this state: for the three largest
concentrations, $\delta_c$ monotonically approaches the arrested state value, as
shown by dashed lines in Fig.~\ref{fig3}(a); by contrast, for $\phi=0.0613$ and
$0.03125$, $\delta_c$ exhibits a pronounced maximum at intermediate times,
followed by a sudden drop towards the arrested state, shown with solid lines in
the figure, which may reflect the formation and subsequent disappearance of a
fluid of particle clusters~\cite{Lu2006}. In this intermediate regime, where the
particles are largely aggregated into nearly close-packed clusters, illustrated
in Fig.~\ref{fig1}(d), the mean distance between clusters is larger at lower
$\phi$. Therefore, percolation occurs only for higher $\delta_c$, as shown in
Fig.~\ref{fig3}(a).

\begin{figure}[b]
\begin{center}
\includegraphics[scale=0.34,clip=true]{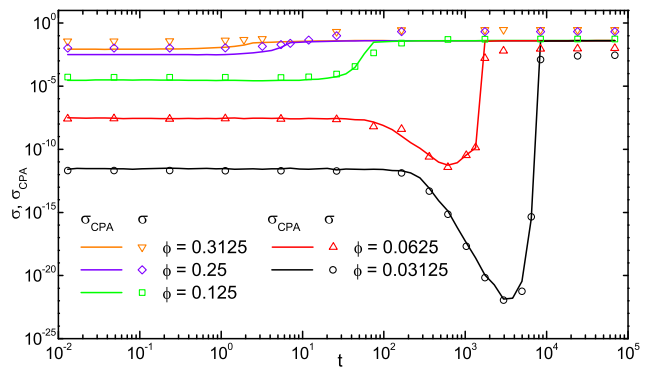}
\caption{
(Color online) Time evolution of conductivity $\sigma$ for $\xi/D=0.1$ during
the formation of the colloidal gel. Numerical solution of the tunneling resistor
equations shown with symbols; $\sigma_\mathrm{CPA}$ obtained from Eq.~\ref{CPA}
using $\sigma_0=0.1$, with solid lines.}\label{fig4}
\end{center}
\end{figure}
To assess the applicability of these simulation predictions to physical systems,
we repeat analysis on gels formed in an experimental attractive
colloid system~\cite{Lu2008}. We use sterically-stabilized
polymethylmethacrylate (PMMA) spheres in a solvent mixture of
decahydronaphtalene and bromocyclohexane~\cite{Lu2006}, with $D\simeq 1120$ nm
and $\phi=0.045$, and introduce a non-adsorbing linear polymer, polystyrene with
molecular weight $M_\mathrm{W}=695,000$, that forms random coils in solution
with radius $R_p=33$ nm, so that $\lambda = 0.06$~\cite{Lu2008}. We select a
sample with polymer concentration $c_\mathrm{p}$=3.31 mg/ml, which phase
separates and arrests to form a gel~\cite{Lu2008}. Using confocal
microscopy~\cite{Lu2007}, we locate each particle
individually~\cite{Lu2006,Lu2007,Lu2008}, thereby allowing the same analysis as
performed on the MD simulation configurations. We observe that the evolution of
experimental $\delta_c$ is in qualitative agreement with the MD simulations at
similar $\phi$, as shown in Fig.~\ref{fig3}(b). Although the initial low-time
plateau cannot be sampled practically in these experiments, a maximum of
$\delta_c$ is discernible at $t \approx 300$ s, followed by a rapid drop of
$\delta_c$ at longer times. For $t\gtrsim 10^4$ s, the system reaches the
arrested gel state, and $\delta_c/D \approx 0.1$, independent of time. This
transition associated with the formation of an arrested gel, consistent with
behavior observed in simulations, is illustrated by the renderings of the
measured particle positions in Fig.~\ref{fig3}(b).

\begin{figure}
\begin{center}
\includegraphics[scale=0.34,clip=true]{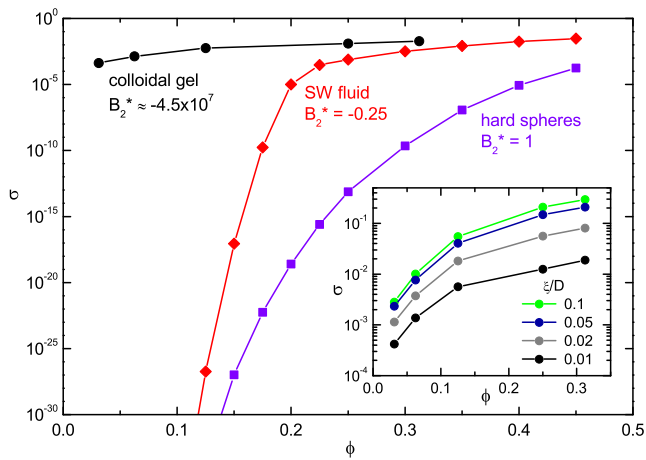}
\caption{
(Color online) Conductivity $\sigma$ as a function of $\phi$ for the arrested
gel state (filled circles), equilibrium SW fluids (filled diamonds), and
equilibrium HS fluids (squares). In all cases, $\xi/D=0.01$. Inset:
$\phi$-dependence of $\sigma$ for the arrested gel state calculated for
different values of $\xi/D$.}\label{fig5}
\end{center}
\end{figure}
We combine the time evolution predictions for $\delta_c$, as shown in
Fig.~\ref{fig3}(a), with Eq.~\eqref{CPA}, to yield an estimate for the time
evolution of $\sigma_\mathrm{cpa}$, and observe that a broad distribution of
$\phi$-dependent $\sigma_\mathrm{cpa}$ conductivities, spanning about ten orders
of magnitude, drastically narrows in the arrested gel state, where
$\sigma_\mathrm{cpa}$ remains at a constant high value for all
$\phi$~\cite{notesigma0}, as shown for $\xi=0.1 D$ with solid lines in
Fig.~\ref{fig4}. Interestingly, for the two lowest $\phi$ values, the maximum of
$\delta_c$ due to the transitory fluid of clusters is reflected by a huge
minimum of $\sigma_\mathrm{cpa}$; fluids of clusters of conducting particles
appear to be substantially worse conductors than a homogeneous fluid of the same
composition.

\begin{figure*}
\begin{center}
\includegraphics[scale=0.4,clip=true]{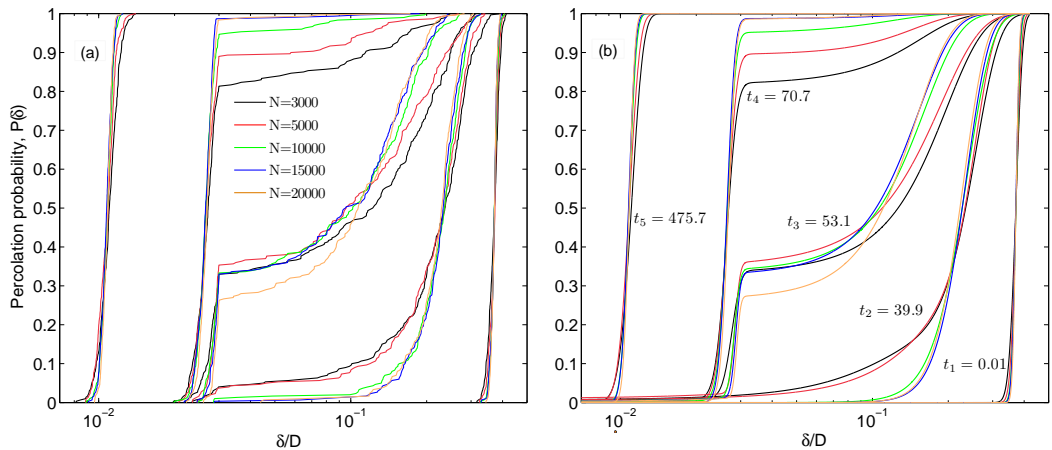}
\caption{
(Color online) (a) Percolation probability $P(\delta)$ as a function of the
connectivity distance $\delta$ at $\phi=0.125$ for different time $t$ (in units
of $D\sqrt{m/u_0}$) and particle numbers $N$. (b) Fits to the results of (a)
using a combination of two error functions.}
\label{fig6}
\end{center}
\end{figure*}
\section{Numerical calculation of the network conductivity}
\label{cond}
To test the accuracy of the results obtained using the CPA, shown
in Fig.~\ref{fig4}, we solve numerically the tunneling resistor network
equations: for each simulation-generated configuration, we assign the
inter-particle conductances from Eq.~\eqref{gij} to each pair of particles,
thereby generating a fully-connected network. To reduce the number of tunneling
connections, we introduce a maximum tunneling distance $\delta_{\rm max}$ so
that the conductances between particles at mutual distances
$\delta_{ij}>\delta_{\rm max}$ are neglected; this does not affect overall
conductivity~\cite{notecond}. For all realizations, we calculate the conductance
$G$ of the reduced network by combining numerical decimation with a
preconditioned conjugate gradient method~\cite{Nigro2012}.
From the dimensionless conductivity $\sigma=GD/L$, where $L$ is the simulation
box edge, we construct the conductivity probability $P(\sigma)$ obtained from
all configurations with fixed $\phi$ and $t$, as shown for tunneling decay
length fixed at $\xi/D = 0.1$ in Figs.~\ref{fig2}(d)-\ref{fig2}(f).

In general, we find qualitative correspondence between $P(\sigma)$ and
$P(\delta)$, which can be seen by comparing Figs.~\ref{fig2}(a)-\ref{fig2}(c)
with Figs.~\ref{fig2}(d)-\ref{fig2}(f).
$P(\sigma)$ and $P(\delta)$ agree even quantitatively for the $\phi=0.03125$,
and for $t$ when the structure is not yet arrested; in this regime
$\delta\gtrsim\xi$, and the CPA provides a good approximation of $\sigma$. These
data confirm the validity of using $P(\sigma)=1/2$ to define the network
conductivity $\sigma$, which we find valid also when $\delta\lesssim\xi$, as
shown in the Appendix. Beginning from the liquid-like states through the onset
of gelation, the $\sigma$ values obtained through this network approach closely
follow the corresponding $\sigma_\mathrm{cpa}$ values, as shown in
Fig.~\ref{fig4} with symbols and lines, respectively. Slight discrepancies in
the $\phi$-dispersion of arrested states likely arise from the short- and
moderately-dispersed distances between the neighboring particles of the spanning
gel structure, which make Eq.~\eqref{CPA} less accurate, as previously
discussed.

\begin{figure}[b]
\begin{center}
\includegraphics[scale=0.33,clip=true]{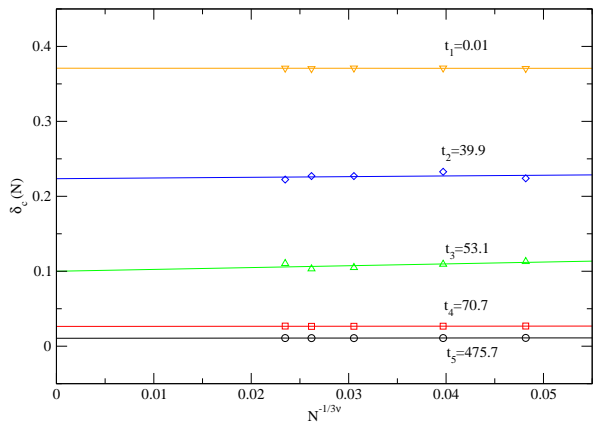}
\caption{
(Color online) Finite-size scaling analysis of the $\delta_c(N)$ values
extracted from $P(\delta_c(N))=1/2$ of Fig.~\ref{fig3}.
Solid lines are fits to Eq.~\eqref{scaling}.}
\label{fig7}
\end{center}
\end{figure}
The significantly higher $\sigma$ in the long-time arrested gel state relative
to the initial fluid-like state, most pronounced for low $\phi$ and highlighted
in Fig.~\ref{fig4}, suggests the general possibility that arrested gel
structures could have higher $\sigma$ relative to other structures formed from
tunneling particles in colloidal suspensions at the same $\phi$. To test this
possibility, we use Monte-Carlo simulations to generate equilibrium fluids of
both HS and SW particles at various $\phi$, with $\lambda=0.03$ and
$\tau=0.2>\tau_c$. For each $\phi$, we obtain $300$ independent equilibrium
configurations of systems with $N=2000$ particles; we determine $\sigma$ for
$\xi/D=0.01$ and compare with the long-time $\sigma$ of the arrested gel state
as a function of $\phi$. In all cases, at any given $\phi$, the gel state has a
higher $\sigma$ than that of the SW fluid, which in turn is always higher than
that of the hard-sphere fluid, as shown in Fig.~\ref{fig5} with circles,
diamonds and squares, respectively. The $\sigma$ values for gel and SW fluid
converge for high $\phi \gg 0.3$; by contrast, for $\phi \lesssim 0.2$, $\sigma$
of the arrested state is many orders of magnitude higher than that of either
fluid. Interestingly, while $\sigma$ depends heavily on $\phi$ in both fluid
cases, it is relatively constant in the gel case, even for $\phi \simeq 0.03$,
as shown in Fig.~\ref{fig5}.

Finally, to explore how the conductivity varies with tunneling decay length, we
calculate $\sigma$ of arrested gels with different $\phi$ and $\xi/D$. We
observe that $\sigma$ only weakly depends on $\xi/D$ in the gel state, due to
short inter-particle distances within the gel, as shown in the inset to
Fig.~\ref{fig5}. Indeed, the relevant length-scale is $\delta_c/D$; in the
arrested state we find $\delta_c\simeq 0.01 D$ and tunneling is thus generally
unaffected so long as $2\delta_c/\xi\lesssim 1$, that is, as long as
$\xi/D\gtrsim 0.02$. At much lower values, $\xi/D$ suppresses inter-particle
tunneling, so that $\sigma$ is small even in the arrested gel state.

\section{Discussion and Conclusions}
\label{concl}
Our data and analyses suggest that, in the arrested regime---where the colloidal
conducting particles form system-spanning amorphous gel structures---the
conductivity can be large and only weakly depends on $\phi$. These results may
impact real-world colloidal systems, where the solid properties of colloidal
gels can be combined with high electrical conductivities to develop materials
with novel mechanical and electrical properties. In contrast to other systems
where conducting particles are embedded in pre-existing gel
networks~\cite{Fizazi1990}, in our conducting colloidal gels the conducting
particles create simultaneously both the gel network and the conducting path.
This synergy can be exploited, at least in principle, to control the
conductivity directly through the attraction between the particles, or to change
the gel structure by local heating of the tunneling-gel network.

\begin{figure*}
\begin{center}
\includegraphics[scale=0.40,clip=true]{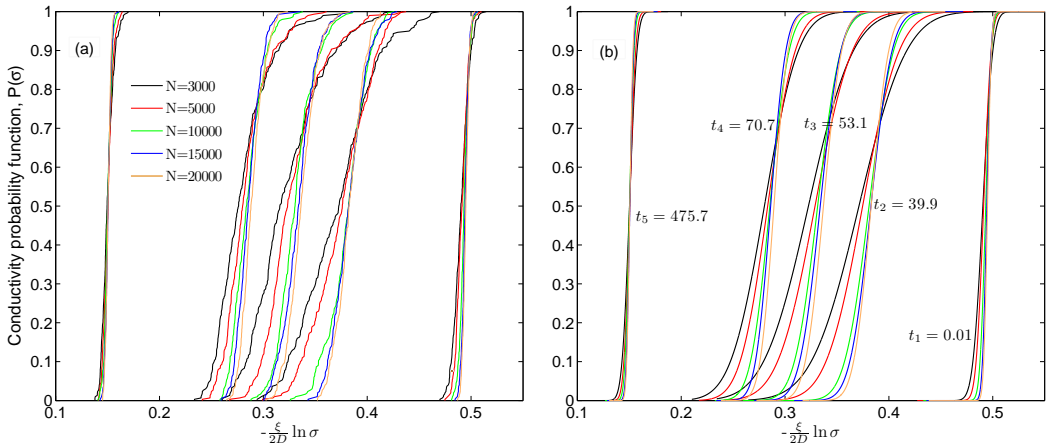}
\caption{
(Color online) (a) Conductivity probability $P(\sigma)$ as a function of
$-\frac{\xi}{2D}\ln(\sigma)$ at $\phi=0.125$ for different values of $t$ (in
units of $D\sqrt{m/u_0}$) and particle numbers $N$. The tunneling decay length
is $\xi=0.1 D$. (b) Fits to the results of (a) using a combination of two error
functions.}
\label{fig8}
\end{center}
\end{figure*}
Our simulation results demonstrate that conduction by tunneling can be strongly
enhanced by the formation of arrested colloidal gel structures; however, the
precise experimental conditions under which systems of real conducting colloidal
systems show similar performance is not yet known. In general, polymers
mediating depletion attractions have gyration radii larger than about $1$-$5$
nm~\cite{Ramakrishnan2002}; therefore, requiring $\xi/D$ to be no more than a
few percent requires $D \gtrsim 50$ nm. Encouragingly, with particles of this
size and typical tunneling decay lengths of a few nm, conductivities like those
in Fig.~\ref{fig5} may yet be achievable in the laboratory. Nevertheless, conducting
particles in this size range remain a significant synthetic challenge, and
suspensions of larger metallic particles show significant sedimentation that may
compromise the formation of gels. Potential solutions to this problem include
using metal-coated PMMA particles, low-structured carbon black particles,
conducting polymer particles, or synthesizing gels in a micro-gravity
environment, such as that provided by the International Space Station.

\acknowledgments B.~N. acknowledges financial support by the Swiss National
Science Foundation (Grant No. 200020-135491). F.~V. and G.~F. acknowledge
financial support by the Swiss National Science Foundation (Grants No.
PP0022-119006 and No. PP00P2-140822/1). 
P.~J.~L. performed the experimental work in the laboratory of
Professor~D.~A.~Weitz at Harvard University, and acknowledges financial support from
NASA (NNX08AE09G, NNX08AE09G S11, NNC08BA08B), the NSF (DMR-1006546), and the
Harvard MRSEC (DMR-0820484).

\appendix
\section{Finite-size analysis of $\delta_c$ and $\sigma$}
We present finite-size scaling analyses of the percolation probability
$P(\delta)$ and of the conductivity probability $P(\sigma)$, demonstrating the validity
of the criteria $P(\delta_c)=1/2$ and $P(\sigma)=1/2$ to define the critical
distance $\delta_c$ and the network conductivity $\sigma$.

\subsection{Critical distance $\delta_c$}
\label{determination}
\begin{table}
\caption{
Finite-size scaling results for $\delta_c$ at $N\rightarrow\infty$ extracted
from the fits shown in Fig.~\ref{fig7} and the corresponding values obtained by
using $P(\delta_c)=1/2$ for $N=10^4$.}
\label{table1}
\begin{ruledtabular}
\begin{tabular}{ccc}
time & $\delta_c(N \rightarrow \infty)$ & $\delta_c(N=10^4)$\\
\hline
$t_1=0.01$ & $0.3709 \pm 0.0007$ & $0.3711$\\
$t_2=39.9$  &  $0.2235\pm 0.0076$  & $0.2270$ \\
$t_3 =53.1$ &  $0.100\pm 0.006$  & $0.1062$ \\
$t_4=70.7$  & $0.0264\pm 0.0003$  & $0.0265$ \\
$t_5=475.7$  &  $0.0106\pm 0.0001$  & $0.0108$
\end{tabular}
\end{ruledtabular}
\end{table}

To demonstrate the finite-size method to determine $\delta_c$ we construct the
percolation probability $P(\delta)$, as described in Sec.~\ref{critical}, for
particle volume fraction fixed at $\phi=0.125$ and for different times $t$ and
particle number $N$, as shown in Figs.~\ref{fig6}(a), with corresponding fits to
$P(\delta)$ obtained from linear combinations of two error functions in
Fig.~\ref{fig6}(b). For the cases with $N>5000$, the fixed point of the
percolation probability---the point at which the $P(\delta)$ curves for
different $N$ intersect each other---is located approximately at $P = 1/2$,
which we use to define critical distance $\delta_c$. To test how this might
change with system size, we apply to each $P(\delta)$ the finite-size scaling
relation:
 \begin{equation}
\label{scaling}
 \delta_c-\delta_c(N)\propto N^{-1/3\nu},
\end{equation}
where $ \nu \simeq 0.88$ is the correlation length exponent and $\delta_c(N)$ is
the value of the critical distance extracted from the fitted curves at exactly
$P(\delta_c(N))=1/2$. From the evolution of $\delta_c(N)$ as a function of
$N^{-1/3\nu}$, we extract $\delta_c$ for $N \rightarrow \infty$ from the
intercept at $N^{-1/3\nu}=0$, shown in Fig.~\ref{fig7}. By comparing these
$\delta_c$ values with those extracted from the condition $P=1/2$ applied to the
$N=10^4$ cases, we find that at worst, $\delta_c(N=10^4)$ is only $6\%$ lower
than its asymptotic estimate at $N\rightarrow\infty$, as seen in
Table~\ref{table1}.

\subsection{Conductivity}
\label{condapp}

\begin{table}
\caption{
Finite-size scaling results for $\sigma$ at $N\rightarrow\infty$, extracted from
the fixed points of $P(\sigma)$ of Fig.~\ref{fig5} and the corresponding values
obtained by using $P(\sigma)=1/2$ for $N=10^4$.}
\label{table2} 
\begin{ruledtabular}
\begin{tabular}{ccc}
time & $-\frac{\xi}{2D} \ln \sigma(N \rightarrow \infty)$ & $-\frac{\xi}{2D} \ln \sigma(N=10^4)$ \\
\hline
$t_1=0.01$  &  $0.4991\pm 0.0006$  & $0.4933$\\
$t_2=39.9$  &  $0.3957\pm 0.0014$  & $0.3809$ \\
$t_3 =53.1$ &  $0.3485\pm 0.0021$  & $0.3314$ \\
$t_4=70.7$  & $0.2957\pm 0.0017$  & $0.2854$ \\
$t_5=475.7$  &  $0.1508\pm 0.0003$  & $0.1503$
\end{tabular}
\end{ruledtabular}
\end{table}
To demonstrate that $P(\sigma)=1/2$ is a valid criterion to extract the network
conductivity $\sigma$, we calculate $P(\sigma)$ for $\xi/D=0.1$, $\phi=0.125$,
and various $N$ and $t$ values, shown in Fig.~\ref{fig8}. We find that the
curves of $P(\sigma)$ do not follow the behavior of the corresponding curves of
$P(\delta)$, as seen by comparing Fig.~\ref{fig8} with Fig.~\ref{fig6}, where
$P(\delta)$ is plotted for the same $N$ and $t$ values. In particular,
$P(\sigma)$ is not affected by the discontinuity of the SW potential, in
contrast to the behavior of $P(\delta)$ when $\delta/D=\lambda$. Furthermore,
the fixed points detected at the crossing of the $P(\sigma)$ as $N$ varies are
no longer associated to $P=1/2$ for $N$ large. However, since the locations of
the fixed points of $P(\sigma)$ are much more precise than those for
$P(\delta)$, they can be used to extract the network conductivity
$\sigma(N\rightarrow\infty)$ with great accuracy. We compare the so-obtained
$\sigma(N\rightarrow\infty)$ values with those obtained from the condition
$P=1/2$ applied to systems with a number of particles fixed at $N=10^4$ in
Table~\ref{table2}; we see that the $P=1/2$ criterion applied to the cases with
$N=10^4$ determines $\sigma$ within $\sim 5$\% of the network conductivity for
$N\rightarrow\infty$.


\begin{thebibliography}{99}

\bibitem{Poon1998} W.~C.~K.~Poon, Curr. Opin. Colloid Interface Sci. \textbf{3}, 593 (1998).

\bibitem{Lekke2002} V.~J.~Anderson and H.~N.~W.~Lekkerkerker, Nature (London) \textbf{416}, 811 (2002)

\bibitem{Sciortino2002} F.~Sciortino, Nat. Mater. \textbf{1}, 145 (2002)

\bibitem{Frenkel2006} D.~Frenkel, Science \textbf{314}, 768 (2006)

\bibitem{Lekke2011}  H.~N.~W.~Lekkerkerker and R.~Tuinier, {\it Colloids and the Depletion Interaction}
(Springer, Dordrecht, 2011).

\bibitem{Manley2004}
S.~Manley et al., Phys. Rev. Lett. \textbf{93}, 108302 (2004).

\bibitem{Lu2008}
P.~J.~Lu, E.~Zaccarelli, F.~Ciulla, A.~B.~Schofield, F.~Sciortino, and
D.~A.~Weitz, Nature (London) \textbf{453}, 499 (2008).

\bibitem{Foffi2005}
G.~Foffi, C.~De Michele, F.~Sciortino, and P.~Tartaglia, Phys.~Rev.~Lett.
\textbf{94}, 078301 (2005).

\bibitem{Zaccarelli2008}
E.~Zaccarelli, P.~J.~Lu, F.~Ciulla, D.~A.~Weitz, and F.~Sciortino,
J.~Phys.~Condens.~Matter \textbf{20}, 494242 (2008).

\bibitem{Lu2006}
P.~J.~Lu, J.~C.~Conrad, H.~M.~Wyss, A.~B.~Schofield, and D.~A.~Weitz,
Phys.~Rev.~Lett. \textbf{96}, 028306 (2006).

\bibitem{Poon2002}
W.~C.~K. Poon, J.~Phys.~Condens.~Matter \textbf{14}, R859 (2002).

\bibitem{Cipelletti2005}
L.~Cipelletti and L.~Ramos, J.~Phys.~Condens.~Matter \textbf{17}, R253 (2005).

\bibitem{Cardinaux2007}
F.~Cardinaux, T.~Gibaud, A.~Stradner, and P.~Schurtenberger,
Phys. Rev. Lett. \textbf{99}, 118301 (2007).

\bibitem{Soga1998}
K.~G.~Soga, J.~R.~Melrose, and R.~C.~Ball, J.~Chem.~Phys. \textbf{108}, 6026
(1998).

\bibitem{FoffiJCP}
G.~Foffi, C.~De Michele, F.~Sciortino and P.~Tartaglia, J. Chem. Phys.
\textbf{122}, 224903 (2005)

\bibitem{Gado2010}
E.~Del Gado, J.~Phys.~Condens.~Matter \textbf{22}, 104117 (2010).

\bibitem{Vigolo2005}
B.~Vigolo, C.~Coulon, M.~Maugey, C.~Zakri, and P.~Poulin, Science \textbf{309},
920 (2005).

\bibitem{Kyrylyuk2011}
A.~V.~Kyrylyuk, M.~C.~Hermant, T.~Schilling, B.~Klumperman, C.~E.~Koning, and
P.~van der Schoot, Nat.~Nanotechnol. \textbf{6}, 364 (2011).

\bibitem{Schilling2007}
T.~Schilling, S.~Jungblut, and M.~A.~Miller, Phys. Rev. Lett. \textbf{98},
108303 (2007).

\bibitem{Nigro2012}
B.~Nigro, C.~Grimaldi, M.~A.~Miller, P.~Ryser, and T.~Schilling,
J. Chem. Phys. \textbf{136}, 164903 (2012).

\bibitem{xivalues}
A.~V.~Nabok, J.~Massey, S.~Buttle, and A.~K.~Ray, IEE Proc.-Circuits Devices
Syst. \textbf{151}, 461 (2004); G.~Sedghi et al., Nat.~Nanotechnol. \textbf{6},
517 (2011).

\bibitem{Noro2000}
M.~G.~Noro and D.~Frenkel, J.~Chem.~Phys. {\bf 113}, 2941 (2000).

\bibitem{Foffi2006}
G.~Foffi and F.~Sciortino, Phys.~Rev.~E \textbf{74}, 050401(R) (2006).

\bibitem{Malijevsky2006}
A.~Malijevsky, S.~B.~Yuste, and A.~Santos, J.~Chem.~Phys. \textbf{125}, 074507
(2006).

\bibitem{Miller2003}
M.~A.~Miller and D.~Frenkel, Phys.~Rev.~Lett. \textbf{90}, 135702 (2003).

\bibitem{Largo2008}
J.~Largo, M.~A.~Miller, and F.~Sciortino, J.~Chem.~Phys. \textbf{128}, 134513
(2008).

\bibitem{Miller2}
M.~A.~Miller, and D.~Frenkel, J.~Chem.~Phys. \textbf{121}, 535
(2004).

\bibitem{Baxter1968}
R.~J.~Baxter, J.~Chem.~Phys. \textbf{49}, 2770 (1968).

\bibitem{Rapaport1997}
D.~C.~Rapaport, \textit{The Art of Computer Simulations}, 2nd ed. (Cambridge
University Press, London, 1997).

\bibitem{CPApapers}
V.~Ambegaokar, B.~I.~Halperin, and J.~S.~Langer, Phys.~Rev.~B \textbf{4},
2612 (1971); M.~Pollak, J.~Non-Cryst.~Solids \textbf{11}, 1 (1972);
B.~I.~Shklovskii and A.~L.~Efros, Sov.~Phys.~JETP-USSR \textbf{33}, 468 (1971);
B.~I.~Shklovskii and A.~L.~Efros, Sov.~Phys.~JETP-USSR \textbf{34}, 435 (1972).

\bibitem{Ambrosetti2010}
G.~Ambrosetti, C.~Grimaldi, I.~Balberg, T.~Maeder, A.~Danani, and P.~Ryser,
Phys.~Rev.~B \textbf{81} , 155434 (2010).

\bibitem{Nigro2011}
B. Nigro, G. Ambrosetti, C. Grimaldi, T. Maeder, and P. Ryser, 
Phys. Rev. B {\bf 83}, 064203 (2011).

\bibitem{Lu2007}
P.~J.~Lu, P.~A.~Sims, H.~Oki, J.~B.~Macarthur, and D.~A.~Weitz, Opt.~Express
\textbf{15} 8702 (2007).

\bibitem{notesigma0}
In plotting the CPA conductivity results of Fig.~\ref{fig4} we have fixed the prefactor appearing in Eq.~\ref{CPA}
to $\sigma_0=0.1$, which is the value found for hard-spheres fluids~\cite{Nigro2012}.

\bibitem{notecond}
For each realization of the system we first calculate the minimum connectivity
length $\delta$ for a given $\phi$ and $t$, as described in Sec.~\ref{critical}.  We then chose
$\delta_{\rm max}$ as to satisfy $\exp(-2\delta_{\rm max}/\xi)=\exp(-M)g(\delta)$, where
$g(\delta)=\exp(-2\delta/\xi)$ and $M=20$, $10$, and $5$
for $g(\delta)<10^{-50}$, $10^{-50}<g(\delta)<10^{-10}$, and $10^{-10}<g(\delta)$, respectively.

\bibitem{Fizazi1990}
A.~Fizazi, J.~Moulton, K.~Pakbaz, S.~D.~D.~V.~Rughooputh, Paul Smith, and
A.~J.~Heeger, Phys.~Rev.~Lett. \textbf{64}, 2180 (1990).

\bibitem{Ramakrishnan2002}
S.~Ramakrishnan, M.~Fuchs, K.~S.~Schweizer, and C.~F.~Zukoski,
J.~Chem.~Phys. \textbf{116}, 2201 (2002);
S.~A.~Shah, Y.-L.~Chen, S.~Ramakrishnan, K.~S.~Schweizer, and C.~F.~Zukoski,
J.~Phys.~Condens.~Matter \textbf{15}, 4751 (2003).


\end{thebibliography}
\end{document}